\documentstyle[aps]{revtex}
\begin{document}
\title{
{\it Ab initio} calculation of excitonic effects in the optical spectra
of semiconductors }

\author{Stefan Albrecht and Lucia Reining}
\address{  Laboratoire des Solides Irradi\'es,
   URA 1380 CNRS -- CEA/CEREM,
   \'Ecole Polytechnique, F-91128 Palaiseau, France }

\author{Rodolfo Del Sole and Giovanni Onida}
\address{
Istituto Nazionale per la Fisica della Materia, Dipartimento di Fisica
   dell'Universit\`a \\ di Roma ``Tor Vergata'',
   Via della Ricerca Scientifica, I--00133 Roma,
   Italy }

\protect{\bigskip}
\bigskip
%\date{\today}
\draft
\maketitle
\bigskip
\begin{abstract}

An {\it ab initio} approach to 
the calculation of excitonic effects
 in the optical  absorption spectra of semiconductors and insulators
 is formulated. It starts from 
a quasiparticle bandstructure calculation and is based  on the relevant
 Bethe--Salpeter equation.
 An application to bulk silicon
 shows a substantial improvement with respect to previous calculations
 in the description of the experimental
 spectrum, for both peak positions and lineshape.

\end{abstract}

%\twocolumn
\section*{}
\narrowtext

Recent advances in {\it ab initio} calculations, mostly Density Functional
Theory -- Local Density Approximation (DFT-LDA) applications, allow to
determine the ground state properties and the Kohn-Sham (KS) electronic
structure\cite{hohk} for even complicated systems. In order to treat excited
states, realistic quasiparticle (QP) energies are then in general obtained
by applying self-energy corrections to the KS energies, usually evaluated in
the $GW$ approximation\cite{hedin}. Excellent agreement of the resulting
bandstructure with experimental data has been found for a wide range of
materials\cite{hyblou,godshsh}. However, spectroscopic properties involving
two-particle excitations are often only poorly described at this
one-particle level. The main example is absorption spectroscopy, where a
simultaneously created electron--hole pair interacts more or less strongly.
As a consequence, in addition to bound exciton states which occur within the
gap, the spectral lineshape above the continuous--absorption edge is
distorted.

The reported qualitative agreement with experiment of many computed KS-LDA
absorption spectra, obtained from one-electron transitions between KS states
\cite{engfa}, is indeed due to a partial cancellation between two principal
errors: namely, the compensation of the large KS-LDA underestimation of the
valence--conduction bandgap, with an overestimation of the absorption onset
induced by calculating the dielectric function entirely within the
one-particle picture. The situation often worsens when only the first error
is corrected by replacing the KS eigenvalues with the realistic QP energies 
\cite{na4,delsgi}. On the other hand, going beyond the one-particle picture
through inclusion of local field and/or exchange--correlation effects within
DFT-LDA in the calculation of the absorption spectrum does generally not
remove the observed discrepancy\cite{bech}. In fact, most of the residual
error stems from the neglect of the electron--hole interaction. 

Up to now, excitonic effects have been rarely calculated from first
principles. Some information about energetic changes can be extracted from
an LDA-based $\Delta$-SCF approach\cite{mauri}. Large excitonic effects on
the spectral properties have been calculated {\it ab initio} in the case of
a small sodium cluster\cite{na4}. This approach has consequently been
generalized with the calculation of the absorption onset for an infinite
system, the Li$_2$O crystal \cite{li2o}. The calculation of the entire
optical spectrum of a solid, finally, still remains a major challenge \cite
{stras}. Quantitatively correct theoretical absorption spectra are indeed
needed as a reference for the interpretation and prediction of experimental
results.

A paradigmatic case is bulk silicon, which is representative for the group
IV, III-V, and II-VI semiconductors. These materials show qualitatively
similar optical spectra, with two major structures at 3--5 eV. The first
peak (E$_1$) has been interpreted as a M$_1$ type critical point transition,
and the second peak (E$_2$) as a M$_2$ type one \cite{wang}. Theoretical
work based on the one-electron approximation, ranging from early empirical
pseudopotential approaches \cite{ppp} to {\it ab initio} DFT-LDA work \cite
{bech}, all yielded the same qualitative result, i.e. an underestimation of
the E$_1$ peak by as much as 50$\%$, reducing it to a weak shoulder of the
generally overestimated E$_2$ peak. In order to go beyond, Louie {\it et al.}
\cite{louie} included local field effects in the calculation of the
dielectric matrix. The resulting spectrum is significantly improved at
higher energies (above 15 eV), but not in the region of interest around 4 eV.

Several authors suggested that strong contributions to the E$_1$ peak could
arise from saddle point excitons \cite{phivel,expe,HS}. Excitonic effects
allowed to explain the measured temperature and pressure dependence of the
lineshape and the symmetry in wavelength modulation reflectance spectra \cite
{expe}. Until now, the most sophisticated calculation of excitonic effects
on the spectral lineshape of silicon was done by Hanke and Sham \cite{HS}.
They performed a semi-empirical LCAO calculation, including local field
effects and the screened electron--hole attraction. As in Ref. \cite{louie},
local field effects alone were shown to transfer oscillator strength to
higher energies and hence to increase the discrepancy with experiment at
lower energies. On the contrary, the electron--hole interaction shifted the
position of the E$_1$ peak to lower energies, and almost doubled its
intensity, while the oscillator strength of the higher energy peaks was
decreased. The overall agreement with experiment was hence improved, and
clear evidence was given for the importance of excitonic effects. However,
the final intensity ratio between the E$_1$ and E$_2$ peaks was reversed, in
disagreement with the experimental spectrum. As pointed out by Wang {\it et
al.} \cite{wang}, the reliability of semi-empirical approaches is limited.
For instance, there are important differences, already at the one-electron
level, between the spectra of Refs. \cite{louie} and \cite{HS}.

In the {\it ab initio} framework, on the other hand, the precision
achievable for the computation of electronic spectra is in general still
poor when compared with the quality of calculated ground state properties.
This work is aimed to shrink this gap, showing how a significant improvement
of the {\it ab initio} calculation of absorption spectra can be obtained.

The absorption spectrum is given by the imaginary part of the macroscopic
dielectric function $\epsilon_M$ 
\begin{equation}
\epsilon_M(\omega ) = 1 - \lim_{{\bf q} \to 0} v({\bf q}) \hat \chi_{{\bf G}%
=0,{\bf G}^{\prime}=0} ({\bf q};\omega ),
\end{equation}
where $\hat \chi ({\bf r},{\bf r}^{\prime};\omega ) = -iS({\bf r},{\bf r},%
{\bf r}^{\prime},{\bf r}^{\prime};\omega )$. $S(1,1^{\prime};2,2^{\prime})$
is the part of the two-particle Green's function which excludes the
disconnected term $-G(1,1^{\prime})G(2,2^{\prime})$, and $G(1,1^{\prime})$
is the one-particle Green's function \cite{HS}. The notation (1,2) stands
for two pairs of space and time coordinates, $({\bf r}_1, t_1;{\bf r}_2,
t_2) $.

Following Ref. \cite{HS}, we start from the Bethe--Salpeter equation for $S,$%
\begin{equation}
S(1,1^{\prime };2,2^{\prime })=S_{0}(1,1^{\prime };2,2^{\prime
})+S_{0}(1,1^{\prime };3,3^{\prime })\Xi (3,3^{\prime };4,4^{\prime
})S(4,4^{\prime };2,2^{\prime }).
\end{equation}
Repeated arguments are integrated over. The term $S_{0}(1,1^{\prime
};2,2^{\prime })=G(1^{\prime },2^{\prime })G(2,1)$ yields the polarization
function of independent quasiparticles $\chi _{0}$, from which the standard
RPA $\epsilon _{M}$ is obtained. The kernel $\Xi $ contains two
contributions: 
\begin{equation}
\Xi (1,1^{\prime },2,2^{\prime })=-i\delta (1,1^{\prime })\delta
(2,2^{\prime })v(1,2)+i\delta (1,2)\delta (1^{\prime },2^{\prime
})W(1,1^{\prime }).
\end{equation}

Considering the first term in the calculation of $S$ is equivalent to the
inclusion of local field effects in the matrix inversion of a standard RPA
calculation. In order to obtain the macroscopic dielectric constant, the
bare Coulomb interaction $v$ contained in this term must, however, be used
without the long range term of vanishing wave vector \cite
{hankedelsolefiorino84}. When spin is not explicitly treated, $v$ gets a
factor of two for singlet excitons. In the second term, $W$ is the screened
Coulomb attraction between electron and hole. It is obtained as a functional
derivative of the self-energy in the $GW$ approximation, neglecting a term $G%
\frac{\delta W}{\delta G}$. This latter term contains information about the
change in screening due to the excitation, and is expected to be small \cite
{strinati}. We limit ourselves to static screening, since dynamical effects in the
electron--hole screening and in the one particle Green's function tend to
cancel each other \cite{bech2}, which suggests to neglect both of them.

In order to solve Eq.\ (2), we have to invert a 4-point function. In Ref. 
\cite{HS} this has been possible due to the use of a very limited basis set.
In an {\it ab initio} plane wave calculation, such a procedure is clearly
prohibitive, when plane waves are chosen as straightforward basis functions.
Instead, the physical picture of interacting electron--hole pairs suggests
to use a basis of LDA Bloch functions, $\psi _{n}({\bf r}),$ expecting that
only a limited number of electron--hole pairs will contribute to each
excitation.

In this basis, $\chi _{0}^{(n_{1},n_{2}),(n_{3},n_{4})}=\delta
_{n_{1},n_{3}}\delta
_{n_{2},n_{4}}(f_{n_{2}}-f_{n_{1}})/(E_{n_{2}}-E_{n_{1}}-\omega )$ and,
after solving for $S$, in the case of static screening, equation (2) can be
written as 
\begin{equation}
S_{(n_{1},n_{2}),(n_{3},n_{4})}=(H_{exc}-I\,\omega
)_{(n_{1},n_{2}),(n_{3},n_{4})}^{-1}(f_{n_{4}}-f_{n_{3}}),
\end{equation}
with 
\begin{eqnarray}
H_{exc}^{(n_{1},n_{2}),(n_{3},n_{4})} &=&(E_{n_{2}}-E_{n_{1}})\delta
_{n_{1},n_{3}}\delta _{n_{2},n_{4}}-i(f_{n_{2}}-f_{n_{1}})\times  \nonumber
\\
&&\int d{\bf r}_{1}\int d{\bf r}_{1}^{\prime }\int d{\bf r}_{2}\int d{\bf r}%
_{2}^{\prime }\ \psi _{n_{1}}({\bf r}_{1})\,\psi _{n_{2}}^{*}({\bf r}%
_{1}^{\prime })\,\Xi ({\bf r}_{1},{\bf r}_{1}^{\prime },{\bf r}_{2},{\bf r}%
_{2}^{\prime })\,\psi _{n_{3}}^{*}({\bf r}_{2})\,\psi _{n_{4}}({\bf r}%
_{2}^{\prime }).
\end{eqnarray}
$I$ is the identity operator. The energies $E_{n}$ are the QP levels.
Together with the above form of $\chi _{0}$ this is consistent with the use
of LDA wavefunctions and updated energy denominators in the Green's function
used to construct the self-energy in the $GW$ calculation. The $f_{n}$ are
Fermi-Dirac occupation numbers. We avoid to invert the matrix $%
(H_{exc}-I\,\omega )$ for each absorption frequency $\omega $ by applying
the identity 
\begin{equation}
(H_{exc}-I\,\omega )^{-1}=\sum_{\lambda ,\lambda ^{\prime }}\frac{|\lambda
>M_{\lambda ,\lambda ^{\prime }}^{-1}<\lambda ^{\prime }|}{(E_{\lambda
}-\omega )},
\end{equation}
which holds for a system of eigenvectors and eigenvalues of a general,
non-hermitian matrix defined by 
\begin{equation}
H_{exc}|\lambda >=E_{\lambda }|\lambda >.
\end{equation}
$M_{\lambda ,\lambda ^{\prime }}$ is the overlap matrix of the (in general
non-orthogonal) eigenstates of $H_{exc}$.

Equation (7) is the effective two-particle Schr\"odinger equation which we
solve by diagonalization. The explicit knowledge of the coupling of the
various two-particle channels, given by the coefficients $%
A_{\lambda}^{(n_1,n_2)}$ of the state $|\lambda>$ in our LDA basis, allows
to identify the character of each transition. (This analysis would be much
more cumbersome if a matrix inversion instead of the spectral representation
was chosen, as in Ref. \cite{HS}.)

The macroscopic dielectric function in Eq. (1) is obtained as 
\begin{eqnarray}
\epsilon _{M}(\omega )=1-\lim_{{\bf q}\to 0}v({\bf q})\sum_{\lambda ,\lambda
^{\prime }}M_{\lambda ,\lambda ^{\prime }}^{-1}
&&\sum_{n_{1},n_{2}}<n_{1}|e^{-i{\bf q\cdot r}}|n_{2}>A_{\lambda
}^{(n_{1},n_{2})}\times  \nonumber \\
&&\sum_{n_{3},n_{4}}<n_{4}|e^{+i{\bf q\cdot r}}|n_{3}>A_{\lambda ^{\prime
}}^{*(n_{3},n_{4})}\frac{(f_{n_{4}}-f_{n_{3}})}{(E_{\lambda }-\omega )}.
\end{eqnarray}
In practice, the KS eigenvalues and eigenfunctions from a DFT-LDA
calculation serve as input to the evaluation of the RPA screened Coulomb
interaction $W$ and the $GW$ self-energy $\Sigma $. The KS eigenfunctions,
together with the QP energies and $W$, are then used in the exciton
calculation. Here each pair of indices ${(n_{1},n_{2})}$ stands for a pair
of bands and one Bloch vector ${\bf k}$ in the Brillouin zone (BZ), since we
are interested in direct transitions only.

In principle, all combinations of bands should be considered. It can,
however, be shown exactly that only pairs containing one filled and one
empty LDA state contribute to (8). Still, the portion of the matrix $H_{exc}$
to be considered is in general non-hermitian, being of the form\cite{ekpa} 
\[
{\bf H} = \left( 
\begin{array}{ccc}
H^{(v_1,c_1),(v_2,c_2)} & H^{(v_1,c_1),(c_2,v_2)} &  \\ 
- H^{*(v_1,c_1),(c_2,v_2)} & -H^{*(v_1,c_1),(v_2,c_2)} &
\end{array}
\right). 
\]
The off-diagonal coupling matrices do not contain the QP transition
energies, but only the interaction elements, which are much smaller in the
case of silicon. Hence, we neglect the latter and separate the Hamiltonian
into two block-diagonal parts: the resonant contributions, which are active
for positive frequencies, and the antiresonant ones, only contributing to
negative frequencies. The matrix of the resonant part by its own is
hermitian, and we therefore obtain the simpler formula 
\begin{equation}
\epsilon _{M}(\omega )=1+\lim_{{\bf q}\to 0}v({\bf q})\sum_{\lambda }\frac{%
\mid \sum_{v,c;{\bf k}}<v,{\bf k}|e^{-i{\bf qr}}|c,{\bf k}>A_{\lambda
}^{(v,c;{\bf k})}\mid ^{2}}{(E_{\lambda }-\omega )}.
\end{equation}

(7) and (9) constitute a set of equations which has been frequently used in
the non-{\it ab initio} framework \cite{strinati,bassani}. Here, it appears
as a particular approximation to the more general formula (8), with
well-defined {\it ab initio} ingredients which are consistent with the $GW$
approach.

We evaluate expression (9) for bulk silicon. The DFT-LDA calculation is
performed using norm-conserving pseudopotentials\cite{BACH}, an energy
cutoff of 15 Ry, and 256 special {\bf k} points in the BZ \cite{monk}. Next, 
$GW$ corrections to the KS band structure are obtained following the
approach of \cite{godneeds}. The quite smooth $GW$ corrections are
interpolated for the denser {\bf k} point mesh needed for the absorption
spectrum. We evaluate equations (7) and (9) using different sets containing
up to 2048 {\bf k} points in the BZ. In order to handle such large matrices,
the symmetry properties of the crystal are exploited. One has to be very
careful in doing so, since the spectrum turns out to be extremely sensitive
to any inconsistency in the phases which may appear when wavefunctions are
rotated, notably for degenerate bands. A safe way to proceed is to make only
partial use of symmetry, considering only those operations which form an
abelian subgroup of the point group, and which altogether allow to
reconstruct the whole zone from a corresponding reduced part. In the
case of silicon, we found it convenient to use the 180$^{0}$ rotations
around the $x$ and the $y$ axis, respectively. These two operations $T$
allow us to break the equation $H^{{\bf kk^{\prime }}}A^{{\bf k^{\prime }}%
}=EA^{{\bf k}}$ (band indices have been suppressed, and repeated indices are
summed over) into four equations to be used for points ${\bf k}_{i}$ in the
reduced zone only. These equations are of the form $h_{{\bf k}_{i}{\bf %
k^{\prime }}_{i}}a^{{\bf k^{\prime }}_{i}}=Ea^{{\bf k}_{i}}$, where $h_{{\bf %
k}_{i}{\bf k^{\prime }}_{i}}=H^{{\bf k}_{i}{\bf k^{\prime }}_{i}}\pm H^{{\bf %
k}_{i}T{\bf k^{\prime }}_{i}}$. The $A^{{\bf k}}$ are then reconstructed
from the reduced eigenvectors $a^{{\bf k}_{i}}$. Moreover, we apply time
reversal and hermiticity in order to accelerate the calculation of the
matrix elements.

A set with 864 {\bf k} points in the full BZ is used to check the various
ingredients of our calculation, in particular the number of bands and the
importance of the off-diagonal elements of the inverse dielectric matrix in
the evaluation of $W$. In the inset of Fig.\ 1, the continuous line shows
the results of a calculation with 4 valence and 4 conduction bands, and the
full $\epsilon ^{-1}$. In the region of interest (below 4.5 eV), a 6 bands
calculation (4 valence + 2 conduction, dotted line) appears to be
sufficient. Neglecting the off-diagonal elements of
 $\epsilon _{{\bf GG}
^{\prime }}^{-1}({\bf q})$ yields an indistinguishable curve. We then use 6
bands and the diagonal $\epsilon ^{-1}$ to compute the spectrum with 2048 
{\bf k} points in the full BZ. In the main part of Fig. 1 the experimental
spectrum (dotted line) \cite{aspnes} is compared to: i) an RPA calculation 
\cite{NOTA} taking only into account the QP shifts, but not the excitonic or
local field effects (short--dashed curve): the result is, as generally
observed, in great discrepancy with experiment; ii) a calculation including
local field effects (i.e. using equation (3) with $W$ set to 0, long--dashed
curve): the agreement is worsened, since the oscillator strength is slightly
shifted to higher energies and both the E$_{1}$ and E$_{2}$ peaks are
lowered, thus confirming previous findings in the literature \cite{louie,HS}%
; iii) finally, the full calculation including the electron--hole attraction
(continuous curve): absolute intensities now agree well with experiment. The
remaining slight overestimate is of the order of what has been predicted by
Ref.\ \cite{bech2} to be the contribution of dynamical effects. More important, 
the peak positions and the relative intensity of the main structures are both 
in good agreement with experiment. Also the structure at 3.8~eV, even though 
slightly overestimated due to a finite ${\bf k}$ point sampling, has 
been repeatedly confirmed in theoretical and experimental work \cite{Cheli76}.

In conclusion, we have shown how excitonic effects can be included in an 
{\it ab initio} calculation of optical absorption spectra of semiconductors.
At the example of bulk silicon, we have demonstrated that good agreement
with experiment can be obtained for a case where the inclusion of
self-energy and local field effects alone still gives rise to a rather poor
theoretical spectrum. In this context, bulk silicon is not particularly easy
to handle, since the bottleneck of the calculation is the number of {\bf k}
points (high in silicon, due to large dispersion) and not the energy cutoff.
Even though, the computational effort, mostly steming from diagonalization,
is reasonable and demands only a few hours on a Cray C98. The present work
opens hence the way to first-principles calculations of optical absorption
spectra with a precision comparable to that typically achieved in ground
state calculations.

We thank F. Bechstedt and R.W. Godby for helpful discussions. This work was
supported in part by the European Community programme ``Human Capital and
Mobility'' through Contract No. ERB CHRX CT930337. Computer time on the Cray
C98 was granted by IDRIS (Project No. CP9/970544).

\narrowtext
\newpage \narrowtext
\begin{figure}[tbp]
\caption{Absorption spectra of Si. Inset: calculation according to equation
(9) with 864 {\bf k} points in the BZ, using 8 bands (continuous curve) or
only 6 bands (dotted curve). Main part: Calculation according to equation
(9) with 2048 {\bf k} points in the BZ, 6 bands and the diagonal
approximation to $\epsilon^{-1}$: with both electron--hole attraction and
local field effects in the Hamiltonian (continuous curve), inclusion of
local field effects alone (long--dashed curve) and RPA with QP shifts only
(short--dashed curve). Experimental curve (dots) [28]. }
\label{fig:spec}
\end{figure}


\begin{references}
\bibitem{stras}  The difficulties and some preliminary results were
illustrated in Ref.\cite{stras2}.
\end{references}
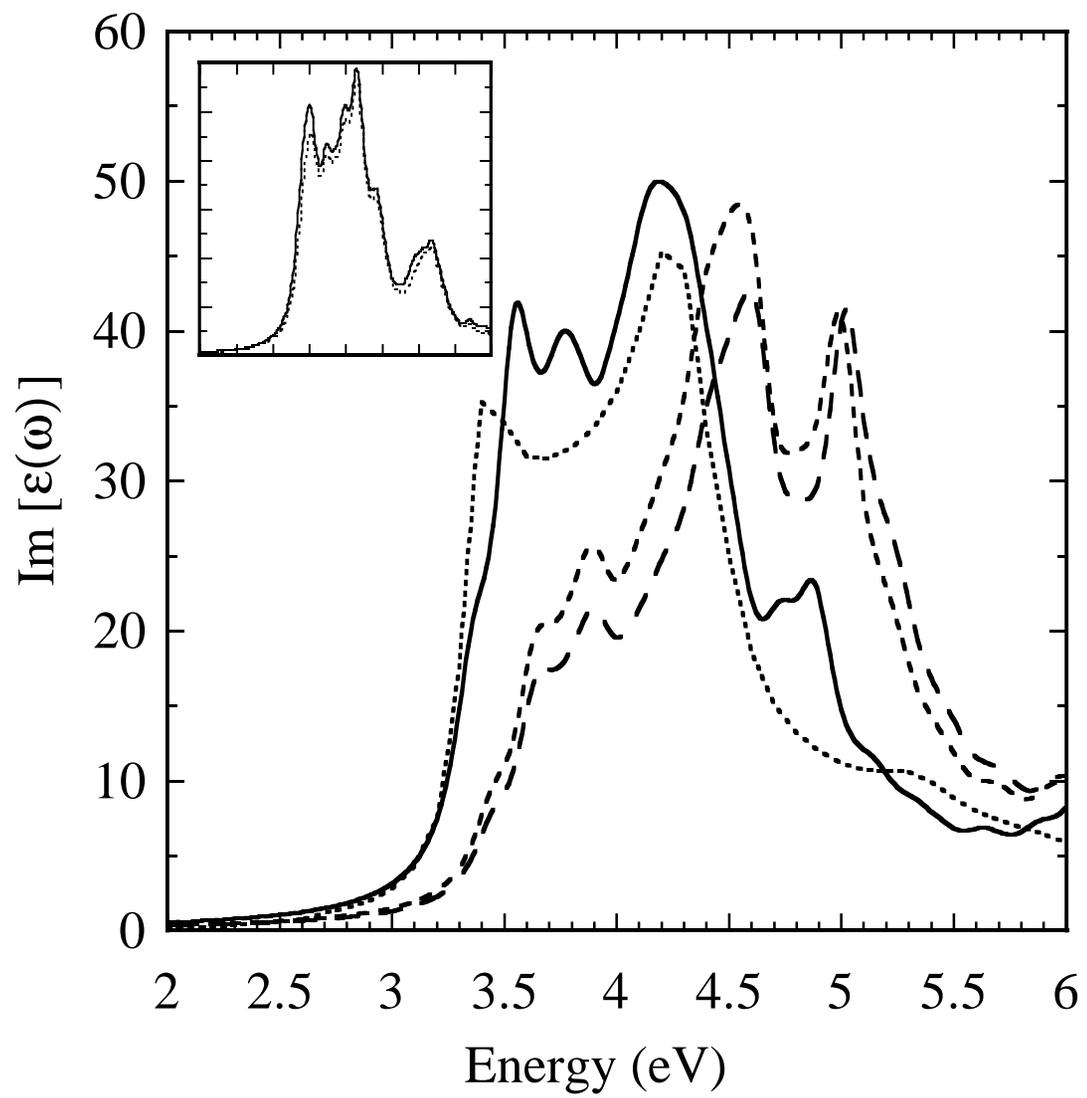
\end{document}